\documentstyle[12pt,rotate,cite]{article}
\newcommand{\be}{\begin{equation}}
\newcommand{\ee}{\end{equation}}
\newcommand{\bea}{\begin{eqnarray}}
\newcommand{\eea}{\end{eqnarray}}
\newcommand{\bd}{\begin{displaymath}}
\newcommand{\ed}{\end{displaymath}}
\newcommand{\f}{\frac}
\newcommand{\ra}{\rightarrow}

\newcommand{\al}{\alpha_s}
\newcommand{\aem}{\alpha}
\newcommand{\Bsee}{$B \ra X_s e^+ e^-$ }
\newcommand{\bsee}{$b \ra s e^+ e^-$ }
\newcommand{\bcenu}{$b \ra c e \bar\nu $ }

\newcommand{\Lms}{\Lambda_{\overline{\rm MS}}}
\newcommand{\Ctilde}{\widetilde C}

\newcommand{\mb}{m_{\rm b}}
\newcommand{\mc}{m_{\rm c}}
\newcommand{\ms}{m_{\rm s}}

\newcommand{\RE}{{\rm Re}}

\begin{document}
\bibliographystyle{physics}
\renewcommand{\thefootnote}{\fnsymbol{footnote}}

\author{
Dongsheng Du${}^{1,2}$~~~ Hongying Jin${}^2$~~~  Yadong Yang${}^{1,2,3}$
\footnote{Email: duds@bepc3.ihep.ac.cn, yangyd@bepc3.ihep.ac.cn}\\
{\small\sl ${}^{1}$ CCAST (World Laboratory), P.O.Box 8730, Beijing 
100080, China}\\
{\small\sl ${}^{2}$ Institute of High Energy Physics, Academia Sinica,
P.O.Box 918(4), Beijing 100039, China\thanks{Mailing address} }\\
{\small\sl ${}^{3}$ Physics Department of Henan Normal University, Xingxiang,
Henan, 453002, China }
}
\date{}
\title{
{\large\sf
\rightline{BIHEP-Th/97-004}
}
\vspace{3cm}
\bigskip
\bigskip
{\LARGE\sf Reexamination of the resonance contributions in
 \Bsee  }
  }
\maketitle
\thispagestyle{empty}
\begin{abstract}
\noindent
With  help of the recent developments in the heavy quarkonium   physics, 
we reexamine the long distance(LD) effects in \Bsee  dominantly from the 
charmonium resonances $J/\Psi$ and $\Psi^{\prime}$ through the decay 
chains $B\ra X_s J/\Psi (\Psi^{\prime}) \ra X_s e^+ e^-$. 
We find that the resonance to nonresonance interference are  
reduced substantially.

\end{abstract}

\newpage
\setcounter{page}{1}

\setcounter{footnote}{0}
\renewcommand{\thefootnote}{\arabic{footnote}}

The rare decay \Bsee has been the subject of many theoretical studies
in the framework of the standard model and its extensions such as the
two Higgs doublet models and models involving supersymmetry
\cite{HWS:87, GSW:89, BBMR:91, AMM:91, JW:90, DPT:93, AGM:94, GIW:94}.
It is believed that once \Bsee
is  observed, it will offer an useful probe of the standard model
and of its extensions. To this end the relevant branching ratio, the
dilepton invariant mass distribution and other distributions of
interest should be calculated with sufficient precision before reliable 
conclusion be drawn. 

At present, the effective
Hamiltonian for \Bsee decays relevant for scale 
$\mu \approx \mb$
in which the short distance QCD effects taken into
account
has been calculated to the next-to-leading order\cite{Mis:93, Mis:94, BM:95} 
in the framework of a renormalization group improved
perturbation theory, and the  
Wilson coefficients in the Hamiltonian are presented to 
have rather small dependence on
$\mu$ and $\Lms$ . 

The actual calculation of \Bsee involves not only the evaluation of
Wilson coefficients of ten local operators  but also the calculation of the
corresponding matrix elements of these operators relevant for
\mbox{\Bsee}, which has been studied in HQET  including  the
non-perturbative ${\cal O}(1/\mb^2)$ corrections  enhanced
 the rate for \Bsee by roughly 4\% \cite{FLS:94, ALI:96}. 
A realistic phenomenological analysis should also
include the long distance contributions which are mainly due to the
$J/\psi$ and $\psi'$ resonances \cite{LMS:89, DTP:89, DT:91, ALI:96, AH:96}.
However,
there are large uncertainties in modeling estimation of such effects and
further study is needed. 
In the papers\cite{LMS:89, DTP:89, DT:91, ALI:96, AH:96},
 the phenomenological parameter  $a_2 $ has to be input in the sub-amplitude 
$A(b\ra s~J/\Psi)$
to compensate  the large discrepancies between theoretical predictions and 
experimental data for $Br(B\rightarrow J/\Psi X_s )$. However, such inputs
would lead to overestimation of the  LD contributions as it will be presented
in what follows.\par
In this paper, we re-examine the LD contributions 
$ B\rightarrow J/\Psi X_s \rightarrow X_s e^+ e^- $ stimulated by the 
recent significant progress in heavy quarkonium physics\cite{BraaY:93, BBL:95}.
We will show that the LD contributions can be substantial reduced when 
the subamplitude  $A(b\ra s J/\Psi )$ decomposed into two
different  parts which do not interfere with each other: color singlet
and color-octet parts as 
in NRQCD \cite{ BBL:95} \par
We will start with the effective Hermitian given as follows \cite{BM:95}
\bea 
{\cal H}_{eff}=-\f{4G_{F}}{\sqrt{2}} V^{*}_{ts}V_{tb}
\sum\limits_{i=1}^{10} C_{i} {\cal O}_{i}
\eea
where the operator basis  is chosen to be 
\be 
\begin{array}{rcl}
Q_1    & = & (\bar{s}_{\alpha}  c_{\beta })_{V-A}
           (\bar{c}_{\beta }  b_{\alpha})_{V-A},       \vspace{0.2cm} \\
Q_2    & = & (\bar{s} c)_{V-A}  (\bar{c} b)_{V-A},     \vspace{0.2cm} \\
Q_3    & = & (\bar{s} b)_{V-A}\sum_q(\bar{q}q)_{V-A},  \vspace{0.2cm} \\
Q_4    & = & (\bar{s}_{\alpha}  b_{\beta })_{V-A} \sum_q (\bar{q}_{\beta}
          q_{\alpha})_{V-A},    \vspace{0.2cm} \\
Q_5    & = & (\bar{s} b)_{V-A}\sum_q(\bar{q}q)_{V+A},  \vspace{0.2cm} \\
Q_6    & = & (\bar{s}_{\alpha}  b_{\beta })_{V-A}
   \sum_q  (\bar{q}_{\beta }  q_{\alpha})_{V+A}, \vspace{0.2cm} \\
Q_7    & = & \f{e}{8\pi^2} m_b \bar{s}_\alpha \sigma^{\mu\nu}
          (1+\gamma_5) b_\alpha F_{\mu\nu},            \vspace{0.2cm} \\
Q_8    & = & \f{g}{8\pi^2} m_b \bar{s}_\alpha \sigma^{\mu\nu}
   (1+\gamma_5)T^a_{\alpha\beta} b_\beta G^a_{\mu\nu}, \vspace{0.2cm} \\
Q_9    & = & (\bar{s} b)_{V-A}  (\bar{e}e)_V ,         \vspace{0.2cm} \\
Q_{10} & = & (\bar{s} b)_{V-A}  (\bar{e}e)_A ,
\end{array}
\ee
where $\alpha$ and $\beta$ denote color indices,  $L$ and $R$ 
denote chiral projections $L(R)=1/2(1\mp\gamma_5 )$.
The Wilson coefficients can be found in \cite{BMMP:94, CFMRS:93,
CFMRS:94, CFRS:94, BM:95, CCRV:94a, CCRV:94b}.

Using the effective Hamiltonian,  the SM-based short distance (SD) 
matrix element for the decay
\bsee can be written as
\bea
{\cal M} (b \ra s e^+ e^- )=\f{G_{F}\alpha}{\sqrt{2} \pi }
V^*_{ts} V_{tb}
 \biggl[
       \left( C_9^{eff} -C_{10} \right) 
(\bar{s}\gamma_{\mu}Lb)(\bar{e}\gamma^{\mu}Le)
 \nonumber\\
+\left( C_9^{eff} +C_{10} \right)
(\bar{s}\gamma_{\mu}Lb)(\bar{e}\gamma^{\mu}Re)
 \nonumber\\
 -2 C_7^{eff}(\bar{s}i\sigma_{\mu\nu}\frac{q^{\nu}}{q^2}m_b Rb)
(\bar{e}\gamma^{\mu}e)\biggr].
\eea
Introducing
\be \label{invleptmass}
\hat s =\f{q^2}{m_b^2 }= \f{(p_{e^+} + p_{e^-})^2}{\mb^2},\hspace{1cm} z =
\f{\mc}{\mb}
\ee
and taking  the spectator approximation, one  finds
\bea \label{rate}
\f{dBr (B \ra X_s e^+ e^- )}{d{\hat s}}
& = &Br (\mbox{\bcenu})\f{\aem^2}{4\pi^2}
\left|\f{V_{ts}}{V_{cb}}\right|^2 \f{(1-\hat s)^2}{f(z)\kappa(z)}
\biggl[(1+2\hat s)\left(|\Ctilde_9^{eff}|^2 + |\Ctilde_{10}|^2\right)
+ \nonumber \\
& & 4 \left( 1 + \f{2}{\hat s}\right) |C_7^{(0)eff}|^2 + 12
C_7^{(0)eff} \ \RE\,\Ctilde_9^{eff}  \biggr],
\eea
\\
where
\bea \label{C9eff}
\Ctilde_9^{eff} & = & \Ctilde_9^{NDR} \tilde\eta(\hat s) + h(z, \hat
s)\left( 3 C_1^{(0)} + C_2^{(0)} + 3 C_3^{(0)} + C_4^{(0)} + 3
C_5^{(0)} + C_6^{(0)} \right) \nonumber \\
& & - \f{1}{2} h(1, \hat s) \left( 4 C_3^{(0)} + 4 C_4^{(0)} + 3
C_5^{(0)} + C_6^{(0)} \right) \nonumber \\
& & - \f{1}{2} h(0, \hat s) \left( C_3^{(0)} + 3 C_4^{(0)} \right) +
\f{2}{9} \left( 3 C_3^{(0)} + C_4^{(0)} + 3 C_5^{(0)} + C_6^{(0)}
\right),
\eea
and 
\bea \label{phasespace}
h(z, \hat s) & = & -\f{8}{9}\ln\f{m_b}{\mu} - \f{8}{9}\ln z +
\f{8}{27} + \f{4}{9} x \\
& & - \f{2}{9} (2+x) |1-x|^{1/2} \left\{
\begin{array}{ll}
\left( \ln\left| \f{\sqrt{1-x} + 1}{\sqrt{1-x} - 1}\right| - i\pi \right), &
\mbox{for } x \equiv \f{4z^2}{\hat s} < 1 \nonumber \\
2 \arctan \f{1}{\sqrt{x-1}}, & \mbox{for } x \equiv \f
{4z^2}{\hat s} > 1,
\end{array}
\right. \\
h(0, \hat s) & = & \f{8}{27} -\f{8}{9} \ln\f{\mb}{\mu} - \f{4}{9} \ln
\hat s + \f{4}{9} i\pi. \\
f(z) & = & 1 - 8 z^2 + 8 z^6 - z^8 - 24 z^4 \ln z, \\
\kappa(z)  & = & 1 - \f{2 \al(\mu)}{3\pi}\left[(\pi^2 - \f{31}{4})(1-z)^2
+ \f{3}{2} \right], \\
\tilde\eta(\hat s) & = & 1 + \f{\al(\mu)}{\pi}\, \omega(\hat s),
\eea
with
\bea \label{omega}
\omega(\hat s) & = & - \f{2}{9} \pi^2 - \f{4}{3}\mbox{Li}_2(\hat s) - \f{2}{3}
\ln {\hat s} \ln(1-{\hat s}) - \f{5+4{\hat s}}{3(1+2{\hat s})}
\ln(1-{\hat s}) - \nonumber \\
& &  \f{2 {\hat s} (1+{\hat s}) (1-2{\hat s})}
{3(1-{\hat s})^2 (1+2{\hat s})} \ln {\hat s} + \f{5+9{\hat s}-6{\hat s}^2}{6
(1-{\hat s}) (1+2{\hat s})}.
\eea
Where $f(z)$ and $\kappa(z)$ are the phase-space factor and the single
gluon QCD correction to the \bcenu decay \cite{CM:79, KM:89}
respectively.  $\tilde\eta$  represents single gluon
corrections to the matrix element of $Q_9$ with $\ms = 0$
\cite{JK:89, Mis:94}.

Now we implement the LD  contributions in \Bsee. The resonance amplitude,
which includes a Breit-Wigner form for the intermediate $\Psi(nS) $ 
state, is given by 
the well known formula 
\bea 
 A(B\ra X_s V \ra X_s e^+ e^- )=
 \f{G_{F}\alpha}{\sqrt{2} \pi }
V^*_{ts} V_{tb}
  (\bar{s}\gamma_{\mu}Lb) ( \bar{e}\gamma^{\mu}e ) \nonumber\\
\times ( C_1 +\f{C_{2}}{3} ) \f{16 \pi^2 }{3} 
\f{f_{V}^2 /m_V^2 }{m_V^2 -s -i\Gamma m_V },
\eea
where   $f_V $ is the decay constant of vector meson V 
$(\Psi, \Psi^{\prime})$ determined by 
\bea 
\Gamma(V\ra e^+ e^- ) =\f{16\pi \alpha^2 }{27m_V^3 }f_{V}^2.
\eea
In the numerical calculations in the  literature
\cite{LMS:89, DTP:89, DT:91, ALI:96, AH:96},
the combination of the Wilson coefficients $C_1 +C_2 /3$ 
is  treated as a phenomenological parameter $a_2$ with a value 0.24
 on the consideration of   the experiment
 data $Br(B\ra X_s J/\Psi)= 0.8\pm 0.08 \% $, 
 which is much larger than the QCD prediction $C_1 +C_2 /3 = 0.12$,
 because the usual calculation 
using  the  Wilson coefficient would produce very low predictions of the 
branching ratio $Br(B\ra X_s J/\Psi)$. However, the  theory of heavy quarkonium 
 production and decays
 has recently undergone  a number of significant developments \cite{BraaY:93, 
 BBL:95}, where color-octet mechanism and factorization have been developed.
It has been shown that the color-octet contributions  compete with the     
color-singlet contributions in the process 
$B\ra X_s J/\Psi(\Psi^{\prime}) $ \cite{KO:96} and 
the predictions agree  with the  experiment data.

Including the color-octet contributions, one can get  
\bea
 A(B\ra X_s \Psi \ra X_s e^+ e^- )=
 \f{G_{F}\alpha}{\sqrt{2} \pi }
V^*_{ts} V_{tb}
  \left[ C^{sig}  
  \langle  X_s^1 |\bar{s}\gamma_{\mu}Lb | B\rangle 
  \langle \Psi |\bar{c}\gamma^{\mu}Lc |X^1 \rangle \right. \nonumber\\
\left.  + C^{oct}  
  \langle  X_s^8 |\bar{s}\gamma_{\mu}LT^a b | B\rangle 
  \langle \Psi |\bar{c}\gamma^{\mu}LT^a c |X^8 \rangle \right] \nonumber\\
\times \bar{e} \gamma^{\mu} e \f{16 \pi^2 }{3} 
\f{f_{\Psi} /m_{\Psi}^2 }{m_{\Psi}^2 s -i\Gamma m_{\Psi}} + (\Psi \ra \Psi^{\prime} ),
 \eea
where $C^{sig}=(C_1 +C_2 /3)$ and $ C^{oct}=2C_2 $. The term proportional 
to $C^{sig}$ is the color singlet amplitude and the term  proportional 
to $C^{oct }$ is the color-octet  amplitude. The color-singlet contribution
to $B\ra X_s l^+ l^- $ can be included directly using eq.(13) as a
modification to $\Ctilde_9^{eff}$ as the usual cases in  
\cite{LMS:89, DTP:89, DT:91, ALI:96, AH:96}. However, the
color-octet contribution can not be treated in such way, because the color-
octet  amplitude has different color structure from the SD  amplitude
in eq.(3)  and does not interfere with it within NRQCD
\cite{BraaY:93, BBL:95} just as the cases in \cite{bod:92,KO:96,flem:97}, 
so, the LD contribution is 
reduced, 
especially, in the low $q^2 $ region. 
 Based on such an  observation, we have 
 \bea
\f{dB_r (b \ra s e^+ e^- )}{d{\hat s}}
& = &B_r (\mbox{\bcenu})\f{\aem^2}{4\pi^2}
\left|\f{V_{ts}}{V_{cb}}\right|^2 \f{(1-\hat s)^2}{f(z)\kappa(z)} 
\biggl[(1+2\hat s)\times \nonumber \\
& &\left(|\Ctilde_9^{eff}+Y^{res}({\hat s}) |^2 +
|\Ctilde_{10}|^2 +\left| C_{oct}({\hat s}) \right|^2 \right)
+ \nonumber \\
& & 4 \left( 1 + \f{2}{\hat s}\right) |C_7^{(0)eff}|^2 + 12
C_7^{(0)eff} \ \RE\,[\Ctilde_9^{eff}+Y^{res}({\hat s})]  \biggr],
\eea
with 
\bea 
Y^{res}({\hat s})
&=& C^{sig} \f{16\pi^2}{3}
\f{f_{\Psi}^2 /m_{\Psi}^2 }{  m_{\Psi}^2- {\hat s}m_b^2
-i\Gamma_{\Psi} m_{\Psi}} + (\Psi \ra \Psi^{\prime}),\\
\left| C_{oct}({\hat s}) \right|^2
&=& C^2_{oct}(\f{16\pi^2}{3})^2 \f{3}{2}
\f{\langle O^{\Psi}_8 ( {}^{3}S_1 ) \rangle }
{\langle O^{\Psi}_1 ( {}^{3}S_1 ) \rangle   }
\f{f_{\Psi}^4 /m_{\Psi}^4 }{({\hat s}m_b^2 - m_{\Psi}^2
)^2  
+\Gamma_{\Psi}^2 m_{\Psi}^2 } + (\Psi \ra \Psi^{\prime})
\eea
where $O^{\Psi,\Psi^{\prime}}_{1,8} ( {}^{3}S_1 ) $ are defined in 
\cite{BBL:95}.
The other colored ${\bar c}c$ pairs produced at short distance with
different quantum number ${^1}S_0$ and ${^3}P_J$ contribute to
$J/\Psi,~\Psi^{\prime}$ productions with  same Wilson coefficients
$C_{oct}$,
\bea 
\Gamma(b\rightarrow J/\Psi +X)\propto 
& &3C^2_{oct}(m^2_b +8m^2_c )
\langle O^{\Psi}_{8} ( {}^{3}S_1 )\rangle \nonumber \\
& &+9C^2_{oct}m^2_b \langle O^{\Psi}_{8} ( {}^{3}S_0 )\rangle \nonumber \\
& &+6C^2_{oct}(m^2_b +8m^2_c )\langle O^{\Psi}_{8} ( {}^{3}P_1 )\rangle.
\eea
The color-octet matrices  
can be calculated by Lattice calculations or fitted out from experiment data.
For simplicity, one can  parameterize color octet contributions  using 
only one matrix $\langle O^{\Psi}_{8} ( {}^{3}S_1 )\rangle$ in eq(18).
Cho and Leibovich \cite{CHO:96} have performed a fit to CDF data including 
the color-octet contribution and found 
\bea 
\langle O^{\Psi}_{8} ( {}^{3}S_1 )\rangle
&=& 1.2\times 10^{-2} GeV^3 ,\\
\langle O^{\Psi^{\prime}}_{8} ( {}^{3}S_1 )\rangle
&=& 7.3\times 10^{-3} GeV^3 .
\eea
Furthermore, using these values,  it  has been shown that theoretical
predictions agree with  experiment results of
$Br(B\rightarrow \Psi X_s )$ and $Br(B\rightarrow \Psi^{\prime} X_s )$
respectively\cite{KO:96,he:97}.
The color-singlet matrix elements determined from leptonic decays of 
$\Psi $ and $ \Psi^{\prime}$ are listed as 
\bea 
\langle O^{\Psi}_{1} ( {}^{3}S_1 )\rangle
&=& 1.32 GeV^3 ,\\
\langle O^{\Psi^{\prime}}_{1} ( {}^{3}S_1 )\rangle
&=& 0.53  GeV^3 .
\eea
The numerical results are presented in Fig.1. The thin solid line 
is the short 
distance contribution, the dotted line is the result including long distance 
contribution as the results in ref\cite{ LMS:89, DTP:89, DT:91,
ALI:96, AH:96}, the dotted-dash 
line is our new   result. We can see that
the LD effects are significantly reduced, 
 especially in the region of theoretical and experimental interesting 
 ${\hat s}<0.31$(corresponding the dilepton invariant mass 
 ${\sqrt s} < 0.9m_{\Psi}$), where short distance physic
might be extracted safely . Here, we recall that there are large uncertainties 
in the color matrix elements. However, the uncertainties would not spoil 
our conclusion {\it if the color-octet mechanism is true and provides solutions to
the large heavy quarkonium production data}.

Note that in  extrapolating the dilepton invariant masses away from the 
resonance region, no extra $q^2$ dependence is included in the 
$V_i -\gamma^* (q^2 )$ junction. (The $q^2$ dependence written explicitly 
in eq.(3) is due to the Breit-Wigner shape of the resonance). 
Introducing  the momentum dependence junction strength  defined by
\bea
\langle 0 \left| {\bar c}\gamma_{\mu}c \right| V(q)\rangle
=f_{V}(q^2 )\epsilon_{\mu}
\eea
and using the formula derived by Terasaki \cite{TERA:81} 
\bea
f_{V}(q^2 )=g_{V}(0)(1+\f{q^2 }{C_V }[d_V -h(q^2 )]),
\eea
where
$C_{\Psi}=0.54,~~C_{\Psi^{\prime}}=0.77,~~d_{\Psi}=d_{\Psi^{\prime}}=0.043$ 
 and  
 \bea
 h({q^2})=\f{1}{16\pi^2 r} \left[-4-\f{20r}{3}+4(1+2r){\sqrt{ \f{1}{r}-1}}
 \arctan\f{1}{{\sqrt {\f{1}{r}-1}}}\right]
 \eea
 with $r=q^2 /m_{V}^2 $ for $0\leq q^2 \leq m_V^2 $. For  $m_V^2 \leq q^2 $, 
 $f_{V}(q^2) =f_{V}(m_V^2 ) $ which is determined by eq(13).
Such a consideration leads to  another considerable
suppression on LD contribution 
in the low dilepton invariant mass  region. 
The numerical results present in the Fig.1 by
a dash line in the region ${\hat s}\leq 0.35$. 

In summary, we have investigated the long distance  effects in 
\Bsee including color-octet contribution. We have shown 
that the LD effects are  reduced substantially  due to the color-octet 
amplitude  not interfering
with the SD amplitude and the color-singlet amplitude.
This result is important for extracting
short distance physics which is a good probe for the physics beyond
the Standard Model. Furthermore, if the momentum dependence
of the $V_i -\gamma^* (q^2 )$ conversion strength is considered,
the long distance effects in the low dilepton invariant mass region
 is negligible  small. The main uncertainty in this paper
 stems from the validity of the application of NRQCD factorization formalism to
charmonium production from B meson decays \cite{bod:92,KO:96,flem:97,he:97}.
 NRQCD factorization formalism is well realized in the studies of high $P_T$
 quarkonium production at hadron colliders, however, in the case of
 B decays, there is very little excess energy left over. In such a situation,
 one should pay attention to the validity. To our knowledge, the
validity of such a particular application of color-octet heavy quarkonium
production ideas has never been justified. Given the validity , another
problem still remains obscure. The soft gluon emitted or absorbed by the
colored ${\bar c}c$ may obscure dilepton  distribution, however, it is
unfortunate that little progress has been made in understanding the role
played by these soft gluons. Anyway,  progress in understanding
$B\rightarrow Charmonium$ decays will improve our predictions on
\Bsee .

\par

\bigskip
\noindent
{\large\bf Acknowledgment}

\noindent
Y.D. Yang thanks C.S.Kim for helpful discussions and the hospitality of ICTP
where this paper is revised. This work is supported 
 in part by the National Natural Science Foundation and the 
Grant of State Commission of Science and Technology of China.

\newpage
\begin{center}
{\large Figure Captions}
\end{center}

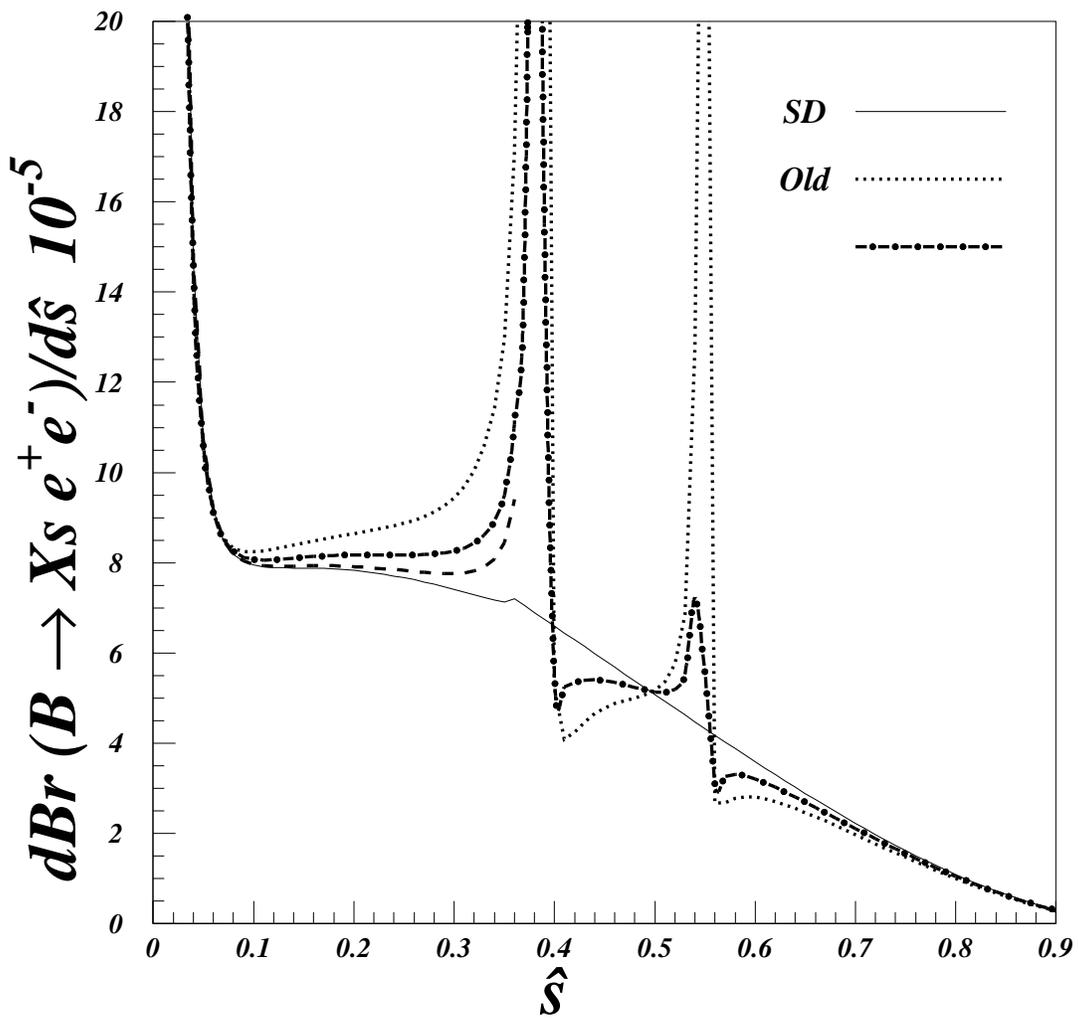
Fig.1. The dileptonic invariant mass spectrum for the decay
\Bsee.
The thin solid and the dotted lines correspond to 
a spectrum without resonances effects, with resonances but
phenomenological parameter $a_2$ used as in\cite{LMS:89, DTP:89, DT:91,
ALI:96, AH:96}. The dotted dash line is the results including the 
color-octet effects. The dash line is  as same as  the dotted-dash line but 
including the momentum dependent $V-\gamma$ conversion strength  effects.
\end{document}